\documentclass{eptcs}

\usepackage{iftex}

\ifpdf
  \usepackage{underscore}         
  \usepackage[T1]{fontenc}        
\else
  \usepackage{breakurl}           
\fi

\usepackage[square,sort,comma,numbers]{natbib}
\usepackage{relsize}
\usepackage[svgnames,dvipsnames]{xcolor}
\usepackage{amssymb}
\usepackage[T1]{fontenc}
\usepackage{mathtools}
\usepackage{breqn}
\newcommand{\sfrac}[2]{\frac{#1}{#2}}
\usepackage{braket}
\usepackage{physics}
\usepackage{tikz}
\usepackage{xparse}
\usepackage{floatrow}
\usepackage{varwidth}
\usepackage[inline]{enumitem}
\usepackage{bm}
\usepackage[margin=1.25in,headheight=13.6pt]{geometry}
\usepackage{tasks}
\usepackage[utf8]{inputenc}

\usepackage{amsthm}

\usepackage[noabbrev,capitalize,nameinlink]{cleveref}
\usepackage{changepage}
\usepackage[normalem]{ulem}
\usepackage{booktabs}
\usepackage{array}
\usepackage{graphicx}
\usepackage{soul}
\usepackage[shortcuts]{extdash}
\usepackage{appendix}
\usepackage{pbox}
\usepackage{sectsty}
\usepackage{microtype}
\usepackage{flafter}

\hypersetup{%
  colorlinks,
  bookmarksnumbered,
  pdfencoding=unicode,
  allcolors=DarkSlateGrey
}

\newcolumntype{L}[1]{>{\raggedright\let\newline\\\arraybackslash\hspace{0pt}}m{#1}}
\newcolumntype{C}[1]{>{\centering\let\newline\\\arraybackslash\hspace{0pt}}m{#1}}
\newcolumntype{M}[1]{>{\centering\let\newline\\\arraybackslash\hspace{0pt}$}m{#1}<{$}}
\newcolumntype{R}[1]{>{\raggedleft\let\newline\\\arraybackslash\hspace{0pt}}m{#1}}

\usepackage{setspace}
\usepackage{eqparbox, etoolbox}
\newlist{rdescription}{description}{1}
\AtBeginEnvironment{rdescription}{%
\setlist[rdescription]{leftmargin =\dimexpr\eqboxwidth{Des}+\labelsep}}%

\SetLabelAlign{parright}{\strut\smash{\parbox[t]\labelwidth{\raggedleft#1}}}

\usetikzlibrary{automata, positioning, decorations.pathmorphing, calc, quotes} 

\tikzset{
    every state/.append style={
        execute at begin node=$,
        execute at end node=$
    },
    initial text = 
}

\usepackage{thmtools}
\usepackage{thm-restate}
\declaretheorem{theorem}
\declaretheorem[sibling=theorem]{lemma,corollary}

\theoremstyle{definition}
\newtheorem{fact}[theorem]{Fact}
\newtheorem{definition}[theorem]{Definition}

\crefname{fact}{fact}{facts}
\Crefname{fact}{Fact}{Facts}

\crefalias{constraint}{equation}
\crefname{constraint}{constraint}{constraints}
\Crefname{constraint}{Constraint}{Constraints}
\creflabelformat{constraint}{#2\textup{(#1)}#3} 
\crefalias{sentence}{equation}
\crefname{sentence}{sentence}{sentences}
\Crefname{sentence}{Sentence}{Sentences}
\creflabelformat{sentence}{#2\textup{(#1)}#3}
\crefalias{expression}{equation}
\crefname{expression}{expression}{expressions}
\Crefname{expression}{Expression}{Expressions}
\creflabelformat{expression}{#2\textup{(#1)}#3}

\NewDocumentEnvironment{delineate}{m}{\textcolor{cyan!70!black!}{> > > > Begin: #1 > > > >}}{\textcolor{red!70!black!}{< < < < End: #1 < < < <}}

\mathchardef\mhyphen="2D

\DeclarePairedDelimiter\paren\lparen\rparen

\newcommand{\oh}[1]{\ensuremath{\mathit{o}\paren*{#1}}}

\makeatletter
\newcommand*{\IfItalicsTF}{%
  \ifx\f@shape\my@test@it
    \expandafter\@firstoftwo
  \else
    \expandafter\@secondoftwo
  \fi
}
\newcommand*{\my@test@it}{it}
\makeatother

\newcommand{\contextsensitivemathrm}[1]{\IfItalicsTF{\mathit{#1}}{\mathrm{#1}}}

\newcommand\machineformat[1]{\ensuremath{\contextsensitivemathrm{#1}}}
\newcommand\langclassformat[1]{\ensuremath{\mathsf{#1}}}

\newcommand{\langformat}[1]{\ensuremath{\mathtt{#1}}}

\NewDocumentCommand{\DTIME}{ o m }{\langclassformat{DTIME\IfValueTF{#1}{\paren[#1]{#2}}{\paren*{#2}}}}

\NewDocumentCommand{\defineautomata}{ m m }{%
    \expandafter\NewDocumentCommand\csname#1fa\endcsname{}{\ensuremath{\machineformat{#1fa}}}%
    \expandafter\NewDocumentCommand\csname rt#1fa\endcsname{}{\ensuremath{\machineformat{rt\mhyphen#1fa}}}%
    \expandafter\NewDocumentCommand\csname o#1fa\endcsname{ o }{\ensuremath{\machineformat{1#1fa\IfValueT{##1}{\paren*{##1}}}}}%
    \expandafter\NewDocumentCommand\csname t#1fa\endcsname{ o }{\ensuremath{\machineformat{2#1fa\IfValueT{##1}{\paren*{##1}}}}}%
    \expandafter\NewDocumentCommand\csname#1tm\endcsname{}{\ensuremath{\machineformat{#1tm}}}%
    \expandafter\NewDocumentCommand\csname #2FA\endcsname{ d<> }{\ensuremath{\langclassformat{#2FA}\IfValueT{##1}{\paren*{##1}}}}%
    \expandafter\NewDocumentCommand\csname RT#2FA\endcsname{ d<> }{\ensuremath{\langclassformat{RT\mhyphen#2FA}\IfValueT{##1}{\paren*{##1}}}}%
    \expandafter\NewDocumentCommand\csname O#2FA\endcsname{ o d<> }{\ensuremath{\langclassformat{1#2FA}\IfValueTF{##1}{\paren*{##1\IfValueT{##2}{, ##2}}}{\IfValueT{##2}{\paren*{##2}}}}}%
    \expandafter\NewDocumentCommand\csname T#2FA\endcsname{ o d<> }{\ensuremath{\langclassformat{2#2FA}\IfValueTF{##1}{\paren*{##1\IfValueT{##2}{, ##2}}}{\IfValueT{##2}{\paren*{##2}}}}}%
}

\defineautomata{d}{D}
\defineautomata{n}{N}
\defineautomata{a}{A}
\defineautomata{p}{P}
\defineautomata{q}{Q}
\defineautomata{qc}{QC}

\newcommand\setneg[1]{\ensuremath{\overline{#1}}}

\newcommand{\PL}{\langformat{PAD}}

\newcommand{\lend}{\ensuremath{\rhd}}
\newcommand{\rend}{\ensuremath{\lhd}}

\newcommand{\reject}{\emph{reject}}

\setlist{itemsep=0pt}

\newlist{observations}{enumerate}{1}
\setlist[observations]{
    label=\arabic{*}.,
    ref=\arabic{*}
}
\crefname{observationsi}{observation}{observations}
\Crefname{observationsi}{Observation}{Observations}

\newlist{differences}{enumerate}{1}
\setlist[differences]{
    label=\arabic{*}.,
    ref=\arabic{*}
}
\crefname{differencesi}{difference}{differences}
\Crefname{differencesi}{Difference}{Differences}

\newlist{strategies}{enumerate}{1}
\setlist[strategies]{
    label=Strategy \arabic{*}:,
    ref=\arabic{*},
    labelwidth=\widthof{Strategy 1:},
    leftmargin=\parindent+\labelwidth+\labelsep
}
\crefname{strategiesi}{strategy}{strategies}
\Crefname{strategiesi}{Strategy}{Strategies}

\newlist{turingenum}{enumerate}{1}
\setlist[turingenum]{
    noitemsep,
    labelsep=.5em,
    leftmargin=1em+\parindent,
    labelwidth=1em,
    label=(\Roman{*}),
    ref=\Roman{*}
}

\crefname{turingenumi}{Stg.}{Stgs.}
\Crefname{turingenumi}{Stage}{Stages}

\SetLabelAlign{Center}{\hss#1\hss}

\newcommand{\narrowfont}[3]{\scalebox{#1}[1.0]{#3}}
\newcommand{\turinglabelformat}[1]{\narrowfont{0.85}{-20}{\scriptsize#1}}
\newlength{\turinglabelgap}

\ExplSyntaxOn
\DeclareExpandableDocumentCommand{\IfNoValueOrEmptyTF}{mmm}
 {
  \IfNoValueTF{#1}{#2}
   {
    \tl_if_empty:nTF {#1} {#2} {#3}
   }
 }
\ExplSyntaxOff

\NewDocumentEnvironment{turing}{ O{} m m }
 {\IfNoValueOrEmptyTF{#1}{\setlength{\turinglabelgap}{0em}}{\setlength{\turinglabelgap}{0.5em}}\small\begin{enumerate}[labelsep=0pt,align=left,parsep=0pt,leftmargin=\widthof{\turinglabelformat{#1}}+\turinglabelgap, 
 listparindent=0pt] 
  \item[]\ignorespaces#3\\[0.5em]
  \begin{turingenum}[
    nosep,
    align=Center,
    labelwidth=\widthof{\turinglabelformat{#1}},
    labelsep=\turinglabelgap,
    leftmargin=0em 
  ]}
 {\unskip\end{turingenum}\end{enumerate}}

\usepackage{crossreftools}
\makeatletter
\newcommand{\optionaldesc}[3]{%
  \phantomsection
#1\protected@edef\@currentlabel{#1}\protected@edef\cref@currentlabel{%
    [#3][\arabic{#3}][\cref@result]%
    #1%
  }\label{#2}%
}
\makeatother

\NewDocumentCommand{\defineturingitem}{ m m }{%
    \expandafter\NewDocumentCommand\csname#1item\endcsname{ o o m }{\IfValueTF{##1}{\item[\turinglabelformat{\optionaldesc{##1}{\IfValueTF{##2}{##2}{stg:##1}}{turingenumi}}]\begin{adjustwidth}{#2}{0pt}\ignorespaces##3\end{adjustwidth}}{\item[]\begin{adjustwidth}{#2}{0pt}\ignorespaces##3\end{adjustwidth}}}%
}

\defineturingitem{t}{0em}
\defineturingitem{tt}{1em}
\defineturingitem{ttt}{2em}
\defineturingitem{tttt}{3em}
\defineturingitem{ttttt}{4em}

\makeatletter
\def\squiggly{\bgroup \markoverwith{\lower3.9\p@\hbox{\sixly \scalebox{1.2}[0.65]{\char58}}}\ULon}
\makeatother

\def\mysout{\leavevmode\bgroup\def\ULthickness{1pt}\ULdepth=-.4ex\ULset}

\newcommand{\stkout}[1]{\begingroup\ifmmode\text{\mysout{\ensuremath{#1}}}\else\mysout{#1}\fi\endgroup}

\newcommand{\utkanworry}[1]{\textcolor{red!45!black!90}{\ifmmode\smash[b]{\squiggly{#1}}\else\squiggly{#1}\fi}}

\renewcommand{\textvisiblespace}[1][.7em]{%
  \makebox[#1]{%
    \kern.07em
    \vrule height.5ex
    \hrulefill
    \vrule height.5ex
    \kern.07em
  }
}

\DeclareMathAlphabet{\mathsl}{\encodingdefault}{\familydefault}{m}{sl}
\SetMathAlphabet{\mathsl}{bold}{\encodingdefault}{\familydefault}{bx}{sl}

\title{Subexponential-Time Quantum Advantage beyond Double-Logarithmic Space}
\author{A. C. Cem Say
\institute{Department of Computer Engineering\\
Bo\u{g}azi\c{c}i University\\
\.{I}stanbul, T\"{u}rkiye}
\email{say@bogazici.edu.tr}
}
\def\titlerunning{Subexponential-Time Quantum Advantage beyond Double-Logarithmic Space}
\def\authorrunning{A. C. Cem Say}

\NewDocumentCommand{\padlang}{ e{_} m }{\ensuremath{\PL_{\IfValueT{#1}{#1}}\paren*{#2}}}

\NewDocumentCommand{\padpair}{ s m }{%
    \IfBooleanTF{#1}%
    {\ensuremath{\paren*{\padlang{#2}, \padlang{\setneg{#2}}}}}%
    {\ensuremath{\paren{\padlang{#2}, \padlang{\setneg{#2}}}}}%
}

\NewDocumentCommand{\branch}{ m o }{\makebox{\textsc{[#1]}\IfValueT{#2}{ Branch #2:}}}
\NewDocumentCommand{\branchpr}{ m o }{\branch{\,Probability: \ensuremath{#1}\,}[#2]}
\NewDocumentCommand{\branchperc}{ m o }{\branch{#1\% prob.}[#2]}

\NewDocumentCommand{\vartextvisiblespace}{ O{.7em} O{.7ex} }{%
  \makebox[#1]{%
    \kern.07em
    \vrule height#2
    \hrulefill
    \vrule height#2
    \kern.07em
  }
}

\NewDocumentCommand{\Qpub}{ e{_} }{\ensuremath{Q_{\textnormal{pub}\IfValueT{#1}{,#1}}}}
\NewDocumentCommand{\Qpri}{ e{_} }{\ensuremath{Q_{\textnormal{pri}\IfValueT{#1}{,#1}}}}
\NewDocumentCommand{\Qcom}{ e{_} }{\ensuremath{Q_{\textnormal{com}\IfValueT{#1}{,#1}}}}
\NewDocumentCommand{\bpub}{ e{_} }{\ensuremath{b_{\textnormal{pub}\IfValueT{#1}{,#1}}}}
\NewDocumentCommand{\bpri}{ e{_} }{\ensuremath{b_{\textnormal{pri}\IfValueT{#1}{,#1}}}}

\makeatletter

\def\@testdef #1#2#3{%
  \def\reserved@a{#3}\expandafter \ifx \csname #1@#2\endcsname
 \reserved@a  \else
\typeout{^^Jlabel #2 changed:^^J%
\meaning\reserved@a^^J%
\expandafter\meaning\csname #1@#2\endcsname^^J}%
\@tempswatrue \fi}

\begin{document}

\maketitle

\begin{abstract}

We prove that subexponential-time quantum Turing machines are superior to their classical counterparts within common space bounds in $\Omega(\log \log n)$.
For that purpose, we define infinitely many sets of ``padded palindromes'' that are distinguished from each other by the precise relationships between the lengths of the palindromic prefixes and the paddings.
We exhibit an infinite family $\mathcal{F}$ of   functions in $(\log n)^{\omega(1)}\cap n^{o(1)}$ such that for every $\tilde{f}_{i}\in\mathcal{F}$, there exists another function $\tilde{f}_{i+1}\in\mathcal{F}$ such that $\tilde{f}_{i+1}(n) \in o(\tilde{f}_{i}(n))$, and each such $\tilde{f}_{i}$ corresponds to a different quantum advantage statement, i.e. a proper inclusion of the form $\mathsf{BPTISP}(2^{O(\tilde{f}_{i}(n))},o(\log \tilde{f}_{i}(n)))\subsetneq \mathsf{BQTISP}(2^{O(\tilde{f}_{i}(n))},o(\log \tilde{f}_{i}(n)))$ for a different pair of subexponential time and sublogarithmic space bounds. One can also obtain quantum advantage statements where the common  space bound is $\Theta(\log \log n)$ and the time bound is ``almost'' quasi-polynomial, i.e., of the form  
$2^{(\log n)^{\Theta(f(n))}}$,   
where   $f(n)\in \omega(1)$ is a function that can be selected to  grow very slowly. 
Our results depend on a technique enabling polynomial-time quantum finite automata to control the amount of padding with very fine asymptotic granularity.

\end{abstract}

\section{Introduction}\label{sec:intro}

Although it is widely believed that quantum computation is intrinsically more powerful than classical computation, this difference in power has been proven unconditionally in only a limited number of scenarios. Celebrated examples like Shor's algorithm for integer factorization \cite{S97} do not constitute such proofs when the underlying computational problem is not known to be impossible for classical computers running under comparable time and space bounds. A theoretical demonstration of \textit{quantum advantage} can be expressed as a proper inclusion in terms of simultaneous time-space complexity classes, i.e. in the form 
    \begin{equation*}
        \mathsf{BPTISP}(t(n),s(n))\subsetneq \mathsf{BQTISP}(t(n),s(n)),
\end{equation*}
where $t(n)$ and $s(n)$ are respectively the time and space complexity functions bounding the resource utilization of the compared classical and quantum machines. In this sense, unconditional quantum advantage proofs have only been obtained for sublogarithmic space bounds until now.  

It is known \cite{DS90} that no probabilistic Turing machine (PTM) running within $o(\log \log n)$ space and polynomial expected time can recognize a nonregular language with bounded error. In 2002, Ambainis and Watrous demonstrated a \textit{constant-space} quantum Turing machine, i.e., a two-way  finite automaton with quantum and classical states (2QCFA), that can recognize the nonregular language
\begin{equation*}
    L_{eq}=\Set{0^n 1^n | n\geq 0} 
\end{equation*}
in polynomial expected time \cite{AW02}, thereby proving 
    \begin{equation}\label{eq:awpoly}
        \mathsf{BPTISP}(n^{O(1)},o(\log \log n))\subsetneq \mathsf{BQTISP}(n^{O(1)},o(\log \log n)).
    \end{equation}

Equation \ref{eq:awpoly} is the only unconditionally proven quantum advantage statement where the common time bound is a  polynomial function, and the memory budgets that it covers are extremely restricted. For larger common space bounds, namely, those in $\Omega(\log \log n) \cap o(\log n)$,  Ambainis and Watrous presented another important result: They provided \cite{AW02} a 2QCFA that recognizes $L_{pal}$, the language of palindromes on the alphabet $\{0,1\}$, in exponential expected time. Since no sublogarithmic-space PTM can recognize $L_{pal}$ regardless of its time budget \cite{DS92}, this lets one conclude 
    \begin{equation}\label{eq:awpal}
        \mathsf{BPTISP}(2^{O(n)},o(\log n))\subsetneq \mathsf{BQTISP}(2^{O(n)},o(\log n)).
    \end{equation}

Recently, Remscrim  showed that one cannot obtain a ``tighter'' quantum advantage result (with an improved common time bound) than Equation \ref{eq:awpal} by using $L_{pal}$ to witness the separation: He proved \cite{R21} that no quantum Turing machine (QTM) running in space $o(\log n)$ and expected time $2^{n^{1-\Omega(1)}}$ can recognize $L_{pal}$ with bounded error. 

In this paper, we prove quantum advantage under simultaneous subexponential ($2^{n^{o(1)}}$) time and $\Omega(\log \log n) \cap o(\log n)$   space bounds. For this purpose, we exhibit an infinite family $\mathcal{F}$ of functions in $(\log n)^{\omega(1)}\cap n^{o(1)}$ such that for every $\tilde{f}_{i}\in\mathcal{F}$, there exists another function $\tilde{f}_{i+1}\in\mathcal{F}$ such that $\tilde{f}_{i+1}(n) \in o(\tilde{f}_{i}(n))$, and each such $\tilde{f}_{i}$ corresponds to a different quantum advantage statement of the form 
\begin{equation}\label{eq:sep}
\mathsf{BPTISP}(2^{O(\tilde{f}_{i}(n))},o(\log \tilde{f}_{i}(n)))\subsetneq \mathsf{BQTISP}(2^{O(\tilde{f}_{i}(n))},o(\log \tilde{f}_{i}(n))) 
\end{equation}
for a different pair of subexponential time and sublogarithmic space bounds. Each member $\tilde{f}_{i}$ of $\mathcal{F}$ corresponds to a language $L_i$ which will be used to witness  the corresponding proper inclusion. Each $L_i$ will be a collection of ``padded palindromes'' \cite{FK94,S26}, that is, each member $w$ of $L_i$ will have a prefix $x$ that contains a member of $L_{pal}$, and the length $|x|$ of the palindrome will be bounded by the value $\tilde{f}_{i}(|w|)$.
Our results depend on a technique enabling polynomial-time 2QCFAs to verify this length relation on their input strings for all members of a rich set of padding functions with fine asymptotic granularity.

The rest of the paper is structured as follows: Section \ref{sec:defs} summarizes the machine definitions and previous work on some 2QCFAs that will be used as subroutines in our constructions. Section \ref{sec:padl} describes the common syntax of the  family $\mathcal{L}$ of languages that will be used in our proofs, and 
shows that no member  of that family associated with the padding function $f$ can be 
recognized by a PTM limited to using  $o(\log \hat f(n))$ space, where $\hat {f}$ is a 
total extension of $f$. We conclude the detailed description of $\mathcal{L}$ and the construction of the 2QCFA $M_i$ for the language $L_i$ witnessing the separation of Equation \ref{eq:sep} for any $\tilde{f}_{i}\in\mathcal{F}$ in Section \ref{sec:quantum}.  
Section \ref{sec:conc} explores the applicability of the technique of Section \ref{sec:quantum} to  different function families and exemplifies tight time bounds that can be coupled with space bounds in $\Theta(\log \log n)$ to obtain novel quantum advantage results.

\section{PTMs, QTMs, and useful 2QCFAs}\label{sec:defs}


We will be comparing the capabilities of PTMs and QTMs operating under the same time and space bounds. The formal machine models underlying our discussion are standard \cite{DS90,W03,AW02,R20}, and the reader is directed to \cite{S26} for a detailed exposition where the definitions are aligned for ease of comparison and inter-model compatibility. The following is a brief overview of the points that are salient for our purposes.

Both the PTM and QTM models are defined to have a read-only input tape that is accessible by  a two-way head which is restricted to the tape area containing the string $\lend w \rend$, where $\lend$ and $\rend$ are special endmarker symbols, and  $w$ is the input string. Both models also have  a classical work tape that is accessed by a two-way read/write head, and a finite set of classical states. The QTM model has additional ``quantum-only'' architectural features, namely, a finite quantum register containing a constant number of qubits and a quantum work tape accessed by a dedicated two-way head, whose position is classical. Each tape cell of the quantum work tape contains a qubit. 

The execution of a PTM or QTM terminates when it reaches a halting (accepting or rejecting) state. Both models involve randomness, and the same machine can produce  different (yes/no) outputs and run for different numbers of steps in response to the same input in different executions. We consider expected (not worst-case) runtimes in our analyses. On the other hand, it is standard to measure space complexity using a worst-case metric.  A QTM is said to \textit{run within space} $s(n)$ if, on any input of length $n$, at most $s(n)$ cells are scanned on both the classical and quantum work tapes during the computation. For PTMs, only the classical work tape cells are considered in the analogous definition. 

A (classical or quantum) Turing machine $M$ is said to \textit{recognize} a language $L$ \textit{with error bound} $\varepsilon\in [0,\frac{1}{2})$ if
\begin{itemize}
    \item for each $w\in L$, $M$ accepts $w$ with probability at least $1-\varepsilon$, and
    \item for each $w\notin L$, $M$ accepts $w$ with probability at most $\varepsilon$.
\end{itemize}
If $M$ has the additional guarantee that it accepts all members of $L$ with probability 1, it is said to \textit{recognize $L$ with negative one-sided bounded error}. 

 $\mathsf{BPTISP}(t(n),s(n))$ denotes the class of languages recognizable with bounded error by PTMs running within space $s(n)$ and in expected time $t(n)$.  For QTMs the analogous class is denoted by $\mathsf{BQTISP}(t(n),s(n))$. 

All quantum algorithms that will be described in this paper are constant-space QTMs which never move their quantum work tape heads. Restricting the QTM definition so that the machine never uses the work tapes yields the 2QCFA model \cite{AW02,R20},\footnote{Although the original 2QCFA definition in \cite{AW02} does not restrict the machines to use computable numbers in their transition specifications, our algorithms will conform to the  definition in \cite{S26}, which requires all transition amplitudes to be algebraic numbers and thereby views 2QCFAs as  QTMs without work tapes.} and removing the finite quantum register as well leaves a two-way deterministic finite automaton (2DFA) \cite{RS59}. The  main technical contribution of this paper is the demonstration of the existence of fast 2QCFAs for tasks related to the recognition of padded palindromes.

Our  constructions in the following sections will involve two ``off-the-shelf'' subroutines that target language recognition problems for which efficient 2QCFA algorithms are already known. Both these algorithms operate with negative one-sided bounded error $\varepsilon$, and can be  ``tuned'' (by  increasing the size of the classical state set) to obtain a machine  whose error bound is at most $\varepsilon$, for any desired positive value of $\varepsilon$.

The first of these algorithms is $M_{eq}$ \cite{R20}, which recognizes the language $L_{eq}$ in polynomial expected time. $M_{eq}$ will serve as the basis of a  subroutine  named \textsc{EqLen} to be employed  to compare the lengths of selected  subsequences of the input with each other.

Let 
 \begin{equation}\label{eq:ri}
r_{i}=
\begin{cases}
    1\#1, & \text{if } i = 1, \\
    1^{i}(\#1^i)^{2^i-1}\$r_{i-1}, & \text{if } i > 1,
\end{cases}
\end{equation}
and define 
\[
L_{ruler}=\{r_{i} \mid i\geq 1\}
\]
on the alphabet $\Set{1,\#,\$}$.\footnote{This is an adaptation of a similar pattern used in \cite{S26}.} The second  2QCFA algorithm we will adapt and use in Section  \ref{sec:quantum} 
is $M_{ruler}$, which recognizes the language $L_{ruler}$ in polynomial time \cite{S26}.\footnote{$M_{ruler}$ itself uses several subroutines based on $M_{eq}$.} 

We will also be making use of a  2QCFA algorithm named $M_{pal}$ \cite{AW02}, which recognizes the language
$L_{pal}$.  $M_{pal}$ also operates with negative one-sided error. As mentioned in Section \ref{sec:intro}, the expected runtime of $M_{pal}$ on inputs of length $n$ is $2^{\Theta(n)}$, where the constant ``hidden'' in the big-$\Theta$ notation depends on the allowed error bound $\varepsilon$. In order to ensure that the overall runtimes of our machines remain subexponential, we will be careful to run $M_{pal}$ in a controlled manner on  a specific substring of the input string. The iterative structure of $M_{pal}$, depicted in Figure \ref{fig:Mpal}, will be useful in this regard. The worst-case time complexities of both the stages in the main loop of $M_{pal}$ are in $O(n)$. One sets the parameter $c_{\varepsilon}$ (by a simple manipulation of the state set and transition function of the machine) to a positive integer determined by the desired value of $\varepsilon$, taking care to ensure that the single-round acceptance probability is much smaller than $25^{-n}$. Importantly, $M_{\mathrm{pal}}$ returns the head to the left
endmarker and resets the quantum state at the end of each iteration.
We also arrange for the quantum state to be reset before each
iteration begins. Thus, we will be able to interleave iterations of
$M_{\mathrm{pal}}$ with another subroutine of the algorithm to be
presented in Section \ref{sec:quantum}.

\begin{figure}[htb!]
    \caption{The high-level structure of 2QCFA $M_{pal}$ \cite{AW02}. $w$ denotes the  input string of length $n$.}
    \label{fig:Mpal} 
    \begin{turing}[(RWwww)]{V}{\textit{  }}
        \titem{
            Repeat ad infinitum:}
        \ttitem{\textbf{The Rejection Test.}}
        \tttitem{\{If $w\notin L_{pal}$, this stage \textit{rejects} the input with probability greater than $25^{-n}$.\}}
        \ttitem{\textbf{The Acceptance Test.}}
        \tttitem{\{This stage \textit{accepts} the input with probability  $2^{-c_{\varepsilon}(n+1)}$.\}}
    \end{turing}
\end{figure}



\section{Padded languages and the weakness of small-space PTMs}\label{sec:padl}



Consider  the ``two-track'' alphabet $\Sigma_{2T}= \Sigma_{\textbf{I}}\times \Sigma_{\textbf{II}}$, where
    \begin{itemize}
        \item $\Sigma_{\textbf{I}}=\{0,1,
        \#,\$,\%,\star\}$, and 
        \item $\Sigma_{\textbf{II}}=\{1,\$,\#\}$.
    \end{itemize}
Any string $w$ on this alphabet can be viewed as consisting of two superposed strings to be denoted  $w_{\textbf{I}}$ and $w_{\textbf{II}}$, each of length  $|w|$, defined on $\Sigma_{\textbf{I}}$ and $\Sigma_{\textbf{II}}$ respectively.

\begin{definition}\label{def:l}
    A language $L\subseteq \Sigma_{2T}^*$ is said to be a \textit{padded language (associated with the padding function $f$)} if  the following conditions hold:
\begin{enumerate}
    \item For all $w\in L$,  
    \begin{itemize}
        \item $w_{\textbf{I}}$ is of the form $x\star \Sigma_{\textbf{I}}^+$, where
         the prefix $x$,  to be denoted by the expression \texttt{pre}$(w)$,
          is in $\{0,1\}^+$, and
        \item $w_{\textbf{II}}\in L_{ruler}$. 
    \end{itemize}
    \item There exists a partial function $f: \mathcal{N \rightarrow \mathcal{N}}$ whose domain is $\operatorname{dom}(f)=\{ |w| \mid w \in L \}$ such that  for all $w\in L$, $|\texttt{pre}(w)|\leq f(|w|)$ and there exists a string $s\in L$ such that $|s|=|w|$ and $|\texttt{pre}(s)|=f(|s|)$.
\end{enumerate}
\end{definition}

\begin{definition}\label{def:padpal}
    A padded language $L$ is said to be a set of \textit{padded palindromes} if the following conditions are satisfied for all $w\in L$:
    \begin{enumerate}
        \item $\texttt{pre}(w)\in L_{pal}$, and
        \item for all $x\in L_{pal}\cap \{0,1\}^{|\texttt{pre}(w)|}$, there exists a string $s\in L$ such that $|s|=|w|$ and $\texttt{pre}(s)=x$.
    \end{enumerate}
\end{definition}


In every member $w$ of a set of padded palindromes, the postfix of $w_{\textbf{I}}$ following $\texttt{pre}(w)$ and  the entire content of  $w_{\textbf{II}}$ can be viewed as serving as ``padding'' to encapsulate the palindrome $\texttt{pre}(w)$.

We will make use of the following result by Freivalds and Karpinski about the limitations of  PTMs:

\begin{theorem}\label{thm:FK}
\cite{FK94} For an alphabet $\Sigma$, let $A, B \subseteq \Sigma^*$ with $A \cap B = \emptyset$. Suppose there is an infinite set $I$ of positive integers and functions $g(m), h(m)$ such that $g(m)$ is a fixed polynomial in $m$, and for each $m \in I$, there is a set $W_m$ of words in $\Sigma^*$ such that:
\begin{enumerate}
    \item $\abs{w} \leq g(m)$ for all $w \in W_m$,
    \item there is a constant $c > 1$ 
    such that $\abs{W_m} \geq c^m$, 
    \item for  every $w, w' \in W_m$ with $w \neq w'$, there are words $u, v \in \Sigma^*$ such that:
    \begin{enumerate}
        \item $\abs{uwv} \leq h(m), \abs{uw'v} \leq h(m)$, and
        \item either
            $uwv \in A$ and $uw'v \in B$, or 
            $uwv \in B$ and $uw'v \in A$.
    \end{enumerate}
\end{enumerate}
Then, if a probabilistic Turing machine with space bound $s(n)$ separates $A$ and $B$, then $s(h(m))$ cannot be in $\oh{\log m}$.
\end{theorem}

We are now ready to demonstrate the weakness of small-space PTMs in dealing with sets of padded palindromes. 

\begin{lemma}\label{lem:ptmcannotpaddedpal}
Suppose that  $L$ is an infinite set of padded palindromes associated with an unbounded  padding function $f$, and  
\(\hat f\) is an
arbitrary total extension of \(f\) that is eventually greater than \(1\).
Then $L$ cannot be recognized by a PTM that runs within space $o(\log \hat f(n))$.
\end{lemma}
\begin{proof}
    
    We note that the task of recognizing $L$ is identical to the task of separating $L$ and $\setneg{L}$. 

Let $I$ be the set $\{m \mid m\in \operatorname{Im}(f)\text{ and }m>1\}$, where $\mathrm{Im}(f)=\{f(n)\mid n\in \mathrm{dom}(f)\}$ is the image of the partial function $f$.
Note that $I$ is infinite, since  $f$ is unbounded.
Let $g(m)=m$. Let $h$ be the function that maps each number $m$ in $I$  to the smallest integer $n$  such that $f(n)= m$.

Our choice of $I$ and Definitions \ref{def:l} and \ref{def:padpal} dictate that for  all $m\in I$ 
and for all strings $p$ of length $m$ in $L_{pal}$,  $L$ has a member $z_p$ of length 
$h(m)$
with $\mathtt{pre}(z_p)=p$. Note that all those strings $z_p$ have the same member of $L_{ruler}$ on their second tracks, and we will use the expression $z_{\textbf{II}}$ to denote that common string.

For  any string $w$ and any $i,j$ such that $i\in\{1,\dots,|w|\}$ and $j\in\{i,\dots,|w|\}$,  let $w[i\cdots j]$ denote the substring of $w$ starting at index 
$i$
 and ending at 
$j$. For each $m\in I$, we will have
\[
W_m=\left\{w \mid  w_{\textbf{I}}\in\{0,1\}^{\lfloor\sfrac{m}{2}\rfloor}, w_{\textbf{II}}=z_{\textbf{II}}\left[1\cdots {\left\lfloor\sfrac{m}{2}\right\rfloor}\right]\right\},
\]
defined on the alphabet $\Sigma_{2T}$.

We instantiate the template of   \Cref{thm:FK} with this  $I$, $g$, $h$ and these $W_m$ sets for each $m$. Condition (1) of that theorem is easily seen to be satisfied. Condition (2) is satisfied by $c=2^{1/4}$.

Finally, for every $m$ and $w, w' \in W_m$, we set
the corresponding $u$ and $v$ in the template of \Cref{thm:FK} as follows.
We let $u$ be the empty string.
Consider the palindrome 
     \[
p_w =
\begin{cases}
    w_{\textbf{I}}w_{\textbf{I}}^R, & \text{if } m\text{ is even,} \\
    w_{\textbf{I}}0w_{\textbf{I}}^R  , & \text{if } m\text{ is odd.} 
\end{cases}
\]
     We know that $L$ contains a string $z_{p_w}$ of length $n=h(m)$ whose first track 
     is of the form  $p_w \star t_w$ (for  $t_w\in \Sigma_{\textbf{I}}^+$).
     We  set $v$ to be the string of length $h(m)-\lfloor\sfrac{m}{2}\rfloor$ where  
     \[
v_{\textbf{I}} =
\begin{cases}
    w_{\textbf{I}}^R\star t_w, & \text{if } m\text{ is even,} \\
    0w_{\textbf{I}}^R  \star t_w, & \text{if } m\text{ is odd,} 
\end{cases}
\]
and
\[v_{\textbf{II}}=z_{\textbf{II}}\left[\left\lfloor\sfrac{m}{2}\right\rfloor+1\cdots h(m)\right].\]
 
Clearly, $uwv \in L$, and $uw'v$, whose designated prefix is not a palindrome, is not in $L$ whenever $w'\neq w$. 

Consider any PTM that recognizes $L$ within space bound $s$. 
Since all the conditions of Theorem \ref{thm:FK} are satisfied, $s(h(m))$ cannot be in $o(\log  m)$, where the asymptotic statement quantifies over all sufficiently large $m\in I$.
Since \(h(m)\in \operatorname{dom}(f)\), we have
\[
\hat f(h(m))=f(h(m))=m .
\]
Moreover, \(h(m)\to\infty\) as \(m\to\infty\) over \(I\). Consequently, if \(s(n)\in o(\log \hat f(n))\), then
\[
s(h(m))\in o(\log \hat f(h(m)))=o(\log m)
\]
as \(m\to\infty\) over \(I\), contradicting Theorem \ref{thm:FK}. Hence \(L\) cannot be recognized within space \(o(\log \hat f(n))\). \end{proof}

\section{The power of subexponential-time 2QCFAs}\label{sec:quantum}

In this section, we will prove the existence of a language family $\mathcal{L}=\Set{L_i}_{i\geq 1}$ where each $L_i$ has the following properties:
\begin{enumerate}
    \item $L_i$ is an infinite set of padded palindromes,
      associated with a padding function $f_i$, such that 
        $f_i(n) \in 2^{\Theta\left((\log n)^{1/2^{i}}\right)}$ over $\operatorname{dom}(f_i)$, and 
    \item  For all $w\in L_i$,  
         $w_{\textbf{I}}=\texttt{pre}(w)\star c_w$, such that
         $|\texttt{pre}(w)|= f_i(|w|)$ and the postfix $c_w\in \Sigma_{\textbf{I}}^+$ (the \textit{proof chain}) is of the form $\pi_1\%\pi_2\cdots\%\pi_{m_w}\%\#^+$, where $m_w>0$ and each $\pi_i\in (\Sigma_{\textbf{I}}-\{0,\star,\%\})^+$ is a ``proof'' of a claim about a relationship between the lengths of a specific collection of subsequences of $w$, as will be described below.
\end{enumerate}

We will describe a machine template that can be used to obtain a corresponding 2QCFA $M_i$ for any given $i$ and any desired error bound $\varepsilon$. Each $L_i$ is defined as the language recognized by   bounded error by the machine $M_i$. Our analysis of $M_i$ will reveal that the $L_i$ have the properties listed above, and that 
\[ L_{i} \in \mathsf{BQTISP}\left(2^{2^{O\left((\log n)^{1/2^{i}}\right)}},O(1)\right)
\]
for each $i$.

Each $M_i$ alternates between  two phases:
\begin{enumerate}
    \item In the \textit{padding control} phase, $M_i$ checks whether  its input string $w$ satisfies the following conditions for a specific function $f_i$:
    \begin{itemize}
        \item $w_{\textbf{I}}$ is of the form $ x \star c_w$, where $x\in \{0,1\}^+$ and $c_w\in ((\Sigma_{\textbf{I}}-\{0,\star,\%\})^+\%)^+\#^+$,  
        \item $w_{\textbf{II}}\in L_{ruler}$, and  
        \item $|x|= f_i(|w|)$.
    \end{itemize}
$M_i$ attempts to validate the equality $|x|= f_i(|w|)$ by performing a sequence of length comparison tasks (which will be described in the following subsections) on $w_{\textbf{II}}$, the proof chain $c_w$ and the prefix $x$. This phase operates with negative one-sided error: If $w$ satisfies all of the conditions listed above, $M_i$ proceeds to the next phase. Otherwise, $M_i$ rejects the input with high probability, and goes on to the next phase with the remaining minuscule probability.
    \item In the \textit{palindrome control} phase, $M_i$ checks whether the prefix $x$ is a palindrome. 
    Unlike the padding control phase, each execution of the palindrome control phase may accept the input  (with some very low probability). This phase rejects the input with some low probability if it detects nonpalindromeness, and proceeds to the next phase with the remaining high probability.
\end{enumerate}

Before we start the detailed description of $M_i$ with the padding control phase, let us 
 introduce some terminology that will be useful when talking about submachines working on  selected subsequences of the input string  \cite{S26}. A \textit{signpost}  is a position on the input tape of a 2QCFA that is ``easily locatable'', in the sense that there exists a 2DFA subroutine that moves the head from its location immediately before the activation of that subroutine to that position. Some examples of signposts are the positions of the left and right endmarkers and the $k$'th occurrence (for any constant $k>0$) of a specific symbol on a given track to the right of an already reached signpost. In some  2QCFA descriptions, the notation $\langle  (p_{S_(},p_{S_)}), \Sigma_S\rangle$ will be used to refer to a subsequence consisting of the members of  a set $\Sigma_S\subseteq \Sigma_{\textbf{I}}$  between the two signposts $(p_{S_(}$ and $p_{S_)})$ in the string $w_{\textbf{I}}$ on the first track of the input $w$. Some commonly used subsequences will have dedicated names, for instance, we have already seen in Definition \ref{def:l} that the name  \texttt{pre}$(w)$ is used for  the sequence $\langle  (p_{l},p_{r}), \{0,1\}\rangle$, where $p_{l}$ and $p_{r}$ are the positions of the left endmarker and the leftmost $\star$ in $w_{\textbf{I}}$, respectively.



The algorithms to be described below  use submachines based on the polynomial-time 2QCFA template $M_{eq}$ mentioned in Section \ref{sec:defs}  to test whether the lengths of specific pairs of input subsequences are equal. Say \cite{S26} describes how  a  required subsequence length comparison task can be performed by a modified machine based on the idea of ``running'' $M_{eq}$ on a virtual tape obtained by converting the members of the subsequences in question to virtual symbols in $M_{eq}$'s alphabet. In our descriptions, we will use the ``subroutine call'' \textsc{EqLen}$_{\varepsilon}(S_1,S_2)$ to indicate a submachine that compares the length of subsequence $S_1$ with that of $S_2$, so that the underlying machine is tuned to have negative one-sided error bound $\varepsilon$. Such an \textsc{EqLen} call ends with the 2QCFA rejecting the input if it detects that $|S_1|\neq|S_2|$, and returns successfully otherwise.

We now proceed to describe the allowed  formats for the segments $\pi_j$ of the proof chain $c_w$, and present dedicated submachines for validating whether such a segment is in the corresponding format with bounded error. 




\subsection{Proofs about asymptotically logarithmic relationships}\label{sec:lo}

Recall the pattern  (Equation \ref{eq:ri}) introduced in Section \ref{sec:defs} to define the language $L_{ruler}$. Every string  in $L_{ruler}$ is of the form $1^+\#\{1,\$,\#\}^+$.  For any such string $s$, its prefix up to but not including the first occurrence of $\#$ will be called the \textit{core} of $s$. Cores will be heavily used subsequences in our algorithms. The partial function $\texttt{core}:\{1,\$,\#\}^* \rightarrow  \{1\}^+$ maps any member of  $L_{ruler}$  to its core. Define the partial function $\mathsf{lo}:\mathcal{N}\rightarrow \mathcal{N}$ so that $\mathsf{lo}(n)=|\texttt{core}(s)|$ for any string $s\in L_{ruler}$ of length $n$.  The following facts were  proven in \cite{S26} about the functions $\mathsf{lo}$ and $\mathsf{lo}^{-1}$:
\begin{itemize}
    \item  $\mathsf{lo}^{-1}(n)=n2^{n+1}-1$ for all $n\geq 1$. 
    \item  As a conclusion,  $\mathsf{lo}^{-1}(n)\in 2^{\Theta(n)}$ and    $\mathsf{lo}(n) \in \Theta(\log n)$ on its domain; in fact, $\mathsf{lo}(n) < \log_2 n$ for all $n\in \operatorname{dom}(\mathsf{lo})$.
\end{itemize}

Note that each member $s$ of $L_{ruler}$ can be viewed as a proof of the claim that $|\texttt{core}(s)|=\mathsf{lo}(|s|)$. We can adapt the machine $M_{ruler}$ (Section \ref{sec:defs}) to obtain a polynomial-time subroutine \textsc{Ruler}$_{\varepsilon}(S)$ that checks whether a specified input subsequence $S$ is in $L_{ruler}$  with desired negative one-sided error $\varepsilon$.
In the same vein as \textsc{EqLen}, the subroutine \textsc{Ruler} causes the overall machine to reject the input if it detects that $S \notin L_{ruler}$, and returns successfully otherwise. 

\subsection{Proofs about  multiplications}\label{sec:mult}

Let $i$ and $j$ be  positive integers and consider the string 
\begin{equation}\label{eq:multpat}
    (\#1^j)^{i}
\end{equation}
on the alphabet $\{1,\#\}$.
We will use the following notation to denote subsequences of any string $\pi$ matching the pattern of Expression (\ref{eq:multpat}):
\begin{itemize}
    \item $\mathtt{m1}(\pi)$: The sequence (of length $i$) of $\#$'s in $\pi$
    \item $\mathtt{m2}(\pi)$: The postfix (of length $j$) of $\pi$ that follows the rightmost $\#$
    \item $\mathtt{pr}(\pi)$: The sequence (of length $i\times j$) of $1$'s in $\pi$
\end{itemize}

Clearly, if $\pi$ is sandwiched between two signposts in the input tape, then all these subsequences of $\pi$ are also easily discernible by a two-way finite automaton. This fact will be used in our algorithms.

We describe a 2QCFA subroutine named \textsc{Mult}$_{\varepsilon}(\pi)$ (Figure \ref{fig:mult}) that checks whether a given substring $\pi$ is of the form in Expression (\ref{eq:multpat}): Stage (A) can be performed deterministically. 
The loop (B) verifies that all the 1-segments separated by $\#$'s have equal length. (The algorithm is simplified by the assumption that it will only be run when $\pi$ is a substring in the first track that is immediately followed by a $\%$ symbol, as dictated by the proof chain format.) For an input of length $n$, the tasks of locating the various signposts take $O(n)$ time. Since the loop (B) iterates $O(n)$ times, and each \textsc{EqLen} call takes polynomial expected time, the entire \textsc{Mult} procedure  runs in polynomial time. If $\pi$ is a correct proof about the multiplication $i\times j$ (i.e. if it is in the precise format described above), \textsc{Mult} returns with success with probability 1. The probability of an incorrect proof not leading to rejection is highest when $\pi$ contains a single ``defect'' that gives just one  \textsc{EqLen} call a chance to catch it, and is therefore bounded by $\varepsilon$.


\begin{figure}[htb!]
    \caption{The submachine \textsc{Mult}$_{\varepsilon}(\pi)$.}
    \label{fig:mult} 
    \begin{turing}[(RW)]{V}{\textit{  }}
        \titem[(A)]{If  $\pi$ is not of the  form 
        $(\#1^+)^{+}$,         \reject.}
        \titem{\{Let the signpost $p_l$ denote the position of the leftmost $\#$ in $\pi$.\}}
        \titem[(B)]{While there exists another $\#$ to the right of $p_l$ in $\pi$, do the following:}
                \ttitem{\{Let the signpost $p_m$ denote the position of the nearest  $\#$ in $\pi$ to the right of $p_l$. Let the signpost $p_r$ denote the position of the nearest  $\#$ in $\pi$ to the right of $p_m$. (If no such $\#$ exists, let $p_r$ denote the  position of the $\%$ symbol delimiting  $\pi$ on the right.)\}}
        \ttitem{Run \textsc{EqLen}$_{\varepsilon}(\langle  (p_l,p_m), \{1\}\rangle,\langle  (p_m,p_r), \{1\}\rangle)$.}
        \ttitem{\{Update $p_l$ to denote the position $p_m$.\}}
    \end{turing}
\end{figure}


\subsection{Proof chains and the function family  $\mathcal{F}$}\label{sec:chains}

We are now ready to describe how the 2QCFA $M_i$ performs padding control on its input string $w$. Consider the submachine \textsc{PadCheck}$_{i,\varepsilon}$ whose pseudocode is presented in Figure \ref{fig:padchhier}. All submachine calls in the figure are tuned to the same error bound $\varepsilon$ that is specified for $M_i$. Stage (A) deterministically scans the string $w_{\textbf{I}}$ in the first track. Stage  (B) employs the 2QCFA subroutine \textsc{Ruler} to verify that $w_{\textbf{II}}\in L_{ruler}$ as required. 

\begin{figure}[htb!]
    \caption{The submachine \textsc{PadCheck}$_{i,\varepsilon}$.  $w$ denotes the input string of length $n$.}
    \label{fig:padchhier} 
    \begin{turing}[(RW)]{V}{\textit{  }}
        \titem[(A)]{\textit{Reject} if  $w_{\textbf{I}}$ is not of the form $x \star \pi_1\%\pi_2\cdots\%\pi_{i+1}\%\#^+$, where $x \in \{0,1\}^+$ and each $\pi_j\in (\Sigma_{\textbf{I}}-\{\star,\%\})^+$.}
        \titem[(B)]{
           Run \textsc{Ruler}$_{\varepsilon}(w_{\textbf{II}})$. }
        \titem[(C)]{\fbox{$|\mathtt{core}(w_{\textbf{II}})|=\mathsf{lo}(|w_{\textbf{II}}|)=\mathsf{lo}(n)\in \Theta(\log n)$}}
        \titem[(D)]{Run \textsc{Mult}$_{\varepsilon}(\pi_1)$.}
        \titem[(E)]{
           Run \textsc{EqLen}$_{\varepsilon}(\mathtt{m1}(\pi_1),\mathtt{m2}(\pi_1))$.}
           \titem[(F)]{
           Run \textsc{EqLen}$_{\varepsilon}(\mathtt{core}(w_{\textbf{II}}),\mathtt{pr}(\pi_1))$.}
        \titem{\fbox{$|\mathtt{m1}(\pi_1)|=|\mathtt{m2}(\pi_1)|\in \Theta\left((\log n)^{1/2}\right)$}}
        \titem{For each integer $j$ from 2 up to and including $i$, do the following:}
\ttitem[(G)]{Run \textsc{Mult}$_{\varepsilon}(\pi_j)$.}
        \ttitem[(H)]{
           Run \textsc{EqLen}$_{\varepsilon}(\mathtt{m1}(\pi_j),\mathtt{m2}(\pi_j))$.}
           \ttitem{
           Run \textsc{EqLen}$_{\varepsilon}(\mathtt{m1}(\pi_{j-1}),\mathtt{pr}(\pi_j))$.}
        \ttitem{\fbox{$|\mathtt{m1}(\pi_{j})|=|\mathtt{m1}(\pi_{j-1})|^{1/2}\in \Theta\left((\log n)^{1/2^j}\right)$}}  
        \titem{
           Run \textsc{Ruler}$_{\varepsilon}(\pi_{i+1})$.}
        \titem[(I)]{
           Run \textsc{EqLen}$_{\varepsilon}(\mathtt{core}(\pi_{i+1}),\mathtt{m1}(\pi_i))$.}
        \titem[(J)]{
           Run \textsc{EqLen}$_{\varepsilon}(x,\pi_{i+1})$.}
        \titem{\fbox{$|x|\in 2^{\Theta\left((\log n)^{1/2^i}\right)}$}}  
        \titem{Move the head to the left endmarker.}
    \end{turing}
\end{figure}

Recall that our 2QCFA subroutines have a small probability of failing to reject illegal patterns during such verification attempts. In the pseudocode, each line enclosed in a box contains a statement (about the length of a subsequence in terms of the overall input length $n$)   that would  hold if no subroutine call up to that point made that kind of ``fail-to-reject'' error. Consider, for instance, the line labeled (C) in Figure \ref{fig:padchhier}. Execution reaches that point with probability at most $\varepsilon$ if $w_{\textbf{II}}\notin L_{ruler}$. In that  unlikely case, the equality  $|\mathtt{core}(w_{\textbf{II}})|=\mathsf{lo}(|w_{\textbf{II}}|)$ (Section \ref{sec:lo}) that is supposed to be checked by Stage (B) can be false. If all the conditions checked by the subroutines are actually satisfied by $w$, then all the boxed statements, included to facilitate the analysis below, would also be true, for example, the relation $|\mathtt{core}(w_{\textbf{II}})|\in \Theta(\log n)$ expressed in line (C) would hold for all such $w$.

In a properly padded input string $w$ to $M_i$, the substrings $\pi_1$ through $\pi_i$ are expected to be proofs of multiplication relations in the format of Expression (\ref{eq:multpat}). In Stages (D) through (F), the \textsc{PadCheck}$_{i,\varepsilon}$ algorithm  employs the \textsc{Mult}$_{\varepsilon}$ and \textsc{EqLen}$_{\varepsilon}$ submachines to verify that 
\[
|\mathtt{core}(w_{\textbf{II}})|=|\mathtt{pr}(\pi_1)|=|\mathtt{m1}(\pi_1)|^2.
\]
It then proceeds to similarly  verify that  $|\mathtt{m1}(\pi_j)|$ equals the square root of  $|\mathtt{m1}(\pi_{j-1})|$, establishing the relation $|\mathtt{m1}(\pi_{j})|\in \Theta\left((\log n)^{1/2^j}\right)$ for all  $j$ up to $i$. The final proof string $\pi_{i+1}$ is supposed to be a member of $L_{ruler}$, and is used (Stage (I)) to relate the length of $\mathtt{m1}(\pi_{i})$ to an exponentially longer subsequence, whose length as a function of $n$ is in $2^{\Theta\left((\log n)^{1/2^i}\right)}$, and is required to equal the length of $\texttt{pre}(w)$ in Stage (J).

For any $i>0$, since \textsc{PadCheck}$_{i,\varepsilon}$ is seen to execute a linearly bounded number of polynomial-time submachine calls on the input string, its expected runtime is polynomially bounded. 

Let us say that a string $w$ is \textit{well-padded for $f_i$}  if  $w_{\textbf{I}}$ is of the form checked by Stage (A) in Figure \ref{fig:padchhier} and all the subsequence length relations checked by all the submachines employed by \textsc{PadCheck}$_{i,\varepsilon}$ hold (i.e. every substring examined by a \textsc{Ruler}$_{\varepsilon}$ submachine is  really a member  of $L_{ruler}$, every purported multiplication proof is correct, and every pair of sequences submitted to \textsc{EqLen}$_{\varepsilon}$ are indeed equal in length) when \textsc{PadCheck}$_{i,\varepsilon}$ is run on $w$.\footnote{Note that we have not defined the function $f_i$ yet. This does not create a problem, since the ``$f_i$'' in the definition of ``well-padded for $f_i$'' is just a label depending on $i$.} It is clear that every input string that is well-padded for $f_i$ leads \textsc{PadCheck}$_{i,\varepsilon}$ to completion with probability 1 without rejection. If a string  is not well-padded for $f_i$, it will be rejected by one of the submachines of \textsc{PadCheck}$_{i,\varepsilon}$ with probability at least $1-\varepsilon$.

\begin{lemma}\label{lem:infpad}
    For all $i\ge1$, there exist infinitely many strings that are well-padded for $f_i$.
\end{lemma}
\begin{proof}
    We describe a procedure that can be used to construct a distinct set $W_l$ of  strings that are well-padded for $f_i$ for each $l\ge2$: One starts by setting  $|\mathtt{m1}(\pi_{i})|$ to $l$ and proceeds by instantiating  all named subsequences in Figure \ref{fig:padchhier} to the unique lengths and corresponding  forms required to satisfy the conditions checked by the submachines of \textsc{PadCheck}$_{i,\varepsilon}$.\footnote{When $|\mathtt{m1}(\pi_{i})|$ is set to $l$, the \textsc{EqLen} call at Stage (H) requires setting $|\mathtt{m2}(\pi_{i})|$ to $l$ as well. The \textsc{Mult} call at Stage (G) then dictates that $\pi_{i}$ should be $(\#1^l)^{l}$, and the string is fleshed out constraint by constraint in this manner.} To see that this chain of assignments can be carried out to completion, let us show that the combined length of the prefix $x \star \pi_1\%\pi_2\cdots\%\pi_{i+1}\%$ of the first track
\[
w_{\textbf{I}}
=x \star \pi_1\%\pi_2\cdots\%\pi_{i+1}\%\#^{|w_{\textbf{I}}|-|x \star \pi_1\%\pi_2\cdots\%\pi_{i+1}\%|}
\]
of the string template $w$   that is produced\footnote{The constraint satisfaction procedure described above fixes only the length and not the content of $x$.} will always be shorter than the length of the second track, so that the constraint    $|w_{\textbf{I}}|=|w_{\textbf{II}}|$ is satisfied for all such $l$.

For all $j$ from $i$ down to and including $1$, the proof string $\pi_j$ expresses the multiplication $l^{2^{i-j}} \times l^{2^{i-j}}$ and has length $(l^{2^{i-j}})^2+l^{2^{i-j}}$, as seen in Expression (\ref{eq:multpat}). Stages (I) and (J) of Figure \ref{fig:padchhier} ensure that both $\pi_{i+1}$ and $x$ have the length of the member of $L_{ruler}$ whose core has length $l$. Using the formula (Section \ref{sec:lo}) for $\mathsf{lo^{-1}}$ to compute $|\pi_{i+1}|$ and $|x|$, and accounting for the delimiters $\star$ and $\%$, we see that the total length of $x \star \pi_1\%\pi_2\cdots\%\pi_{i+1}\%$ is
\begin{equation}\label{eq:TOPTR}
2(l2^{l+1}-1)+\sum_{j=1}^i((l^{2^{i-j}})^2+l^{2^{i-j}})+i+2.
\end{equation}

Set $q=l^{2^i}$. Since $|\mathtt{pr}(\pi_1)|=q$, Stages (F) and (B) force the length of the bottom track to be
\[
|w_{\textbf{II}}|=\mathsf{lo^{-1}}(l^{2^{i}})=l^{2^{i}}2^{l^{2^{i}}+1}-1.
\]

Considering the first term in Expression (\ref{eq:TOPTR}),
\[
2(l2^{l+1}-1)<q2^q,
\]
because $q\geq l^2\geq l+2$.

Moreover, for every $j\in \{1,\ldots,i\}$,
\[
l^{2^{i-j}}\leq \bigl(l^{2^{i-j}}\bigr)^2\leq q,
\]
and hence
\[
\sum_{j=1}^i
\left(
\bigl(l^{2^{i-j}}\bigr)^2+l^{2^{i-j}}
\right)
\leq 2iq.
\]
 Finally, $i+2\leq q$ and
$2i+1\leq 2^q-1$, so
\[
2iq+i+2
\leq q(2i+1)
\leq q(2^q-1)
\leq q2^q-1.
\]
It follows that
\[
2(l2^{l+1}-1)
+\sum_{j=1}^i
\left(
\bigl(l^{2^{i-j}}\bigr)^2+l^{2^{i-j}}
\right)
+i+2
<q2^q+(q2^q-1)
=q2^{q+1}-1
=|w_{\mathbf{II}}|,
\]
meaning that the prefix and all the proof substrings fit comfortably in the first track of the template $w$.

The set $W_l$ is then obtained by instantiating $x$ with every possible value from $\Set{0,1}^{|\mathtt{pre}(w)|}$. Every member of $W_l$ is therefore well-padded for $f_i$. Moreover,
all members of $W_l$ have length
\[
\lambda_i(l)=\mathsf{lo}^{-1}(l^{2^i}),
\]
and $\lambda_i(l)$ is strictly increasing with $l$. Thus the nonempty
sets $W_l$, for $l\geq2$, occur at distinct input lengths, proving that
there are infinitely many strings that are well-padded for $f_i$.
\end{proof}

We next note that the prefix length of a well-padded string is uniquely
determined by its total length. Indeed, let $w$ be well-padded for $f_i$,
let $n=|w|$, and put $l=|\mathtt{m1}(\pi_i)|$. The conditions checked by
\textsc{PadCheck}$_{i,\varepsilon}$ imply
\[
\mathsf{lo}(n)
=
|\mathtt{core}(w_{\mathbf{II}})|
=
|\mathtt{pr}(\pi_1)|
=
l^{2^i}.
\]
Consequently,
\[
l=\bigl(\mathsf{lo}(n)\bigr)^{1/2^i},
\]
and Stages~(I) and~(J) imply
\[
|\mathtt{pre}(w)|
=
|\pi_{i+1}|
=
\mathsf{lo}^{-1}\!\left(
\bigl(\mathsf{lo}(n)\bigr)^{1/2^i}
\right).
\]
Thus any two well-padded strings of the same length have prefixes of
the same length. 



We now define $f_i(n)$ as the partial function that maps the length $n$ of any string $w$  that is well-padded for $f_i$  to the number $|\texttt{pre}(w)|$. 
By construction, the set of strings that are well-padded for $f_i$ is a padded language associated with the padding function $f_i$.

For every $f_i$, we define the 
 total extension $\tilde f_i:\mathcal{N \rightarrow \mathcal{N}}$ of $f_i$ to be
\[
\tilde f_i(n)=
\begin{cases}
f_i(n) & \text{if } n\in \mathrm{dom}(f_i),\\
0 & \text{if } n=0, \\
\left\lceil 2^{(\log n)^{1/2^{i}}} \right\rceil & \text{otherwise.}
\end{cases}
\]
(All logarithms are to the base 2.) These total extensions will be useful in expressing  time and space complexities and asymptotic comparisons.

\begin{lemma}
    For each $i>0$, $\tilde f_i(n) \in 2^{\Theta\left((\log n)^{1/2^{i}}\right)}$.  
\end{lemma}
\begin{proof}
By the preceding analysis,    $ f_i(n) \in 2^{\Theta\left((\log n)^{1/2^{i}}\right)}$, i.e., there exist constants $a,b>0$ such that  eventually
\[
2^{a(\log n)^{1/2^{i}}} \leq f_i(n) \leq 2^{b(\log n)^{1/2^{i}}}
\]
on $\operatorname{dom}(f_i)$. For all sufficiently large $n$ outside the domain, we have
\[
 2^{(\log n)^{1/2^{i}}}
\leq
\left\lceil 2^{(\log n)^{1/2^{i}}} \right\rceil
\leq
 2^{(\log n)^{1/2^{i}}}+1
 \leq
  2^{(\log n)^{1/2^{i}}+1}
 \leq
   2^{2(\log n)^{1/2^{i}}}.
\]
Define $a'=\operatorname{min}(a,1)$ and $b'=\operatorname{max}(b,2)$. We see that  
\[
2^{a'(\log n)^{1/2^{i}}} \leq \tilde f_i(n) \leq 2^{b'(\log n)^{1/2^{i}}}
\]
for all sufficiently large $n$,  which is what was to be proved.
\end{proof}

 The function family $\mathcal{F}$ is defined to be the set $\Set{\tilde f_i}_{i\geq 1}$. We note the following facts about $\mathcal{F}$.

\begin{fact}\label{fact}
    For each $i\geq1$, 
    \begin{enumerate}
        \item  $\tilde f_{i+1}(n) \in o(\tilde f_i(n))$, and
\item $\tilde f_{i}(n) \in (\log n)^{\omega(1)}\cap n^{o(1)}$.        
    \end{enumerate} 
        \end{fact}
 
\subsection{Checking padded palindromes
}\label{sec:Mi}

Figure \ref{fig:ana} depicts the     template that will be used to construct a 2QCFA $M_i$ that recognizes a set of padded palindromes associated with padding function $f_i$ within error bound $\varepsilon$.  
Each iteration of the loop starts with the execution of the polynomial-time algorithm  presented in Section \ref{sec:chains} that checks whether the input string $w$ is well-padded for $f_i$, and hence that $|\mathtt{pre}(w)|=f_i(|w|)$, with negative one-sided error $\varepsilon$. 
The linear-time \textsc{PalIter} submachine is obtained   by adapting the code for a \textit{single iteration} of the main loop of the machine $M_{pal}$ (Section \ref{sec:defs}, Figure \ref{fig:Mpal})  to run on  a specified prefix of the input. Recall that both the runtime of $M_{pal}$ and  the number of times that \textsc{PalIter} would need to be iterated to realize that computation on a string of length $m$ are exponential functions of $m$, and we can tune the error bound $\varepsilon$ by manipulating the code of \textsc{PalIter} to set an integer parameter $c_{\varepsilon}$   \cite{AW02}.  

\begin{figure}[htb!]
    \caption{The 2QCFA $M_i$. $w$ denotes the input string. $\varepsilon$ is the error bound.}
    \label{fig:ana} 
    \begin{turing}[]{V}{\textit{  }}
        \titem{
            Repeat ad infinitum:}
        \ttitem{Run \textsc{PadCheck}$_{i,\varepsilon}$.}
        \tttitem{\{This stage 
        \textit{rejects} with probability $1-\varepsilon$ if $w$ is not well-padded for $f_i$.\}}
        \ttitem{
            Run \textsc{PalIter}$_{\varepsilon}(\texttt{pre}(w))$.}
        \tttitem{\{This stage \textit{rejects} with probability greater than $25^{-|\texttt{pre}(w)|}$ if $\texttt{pre}(w)$ is not a palindrome. If it has not rejected, it \textit{accepts} with probability $2^{-c_{\varepsilon}(|\texttt{pre}(w)|+1)}$.\}}
    \end{turing}
\end{figure}

Let us analyze the behavior of $M_i$ in response to  a given input string $w$ of length $n$.

    If $w$ is not well-padded for $f_i$, the very first  run of the padding control phase (and so $M_i$) correctly rejects $w$ with probability at least $1-\varepsilon$.
 If $w$ \emph{is} well-padded for $f_i$, all executions of the padding control phase will be completed without rejection, and  the accept/reject response of $M_i$ will be identical to that of $M_{pal}$ when running on $\texttt{pre}(w)$. Therefore, $M_i$ will accept $w$ with probability 1 if $\texttt{pre}(w)$ is a palindrome, and reject it with probability at least $1-\varepsilon$ otherwise.
 
 As shown in Section \ref{sec:chains}, there are infinitely many strings $w$ that are well-padded for $f_i$. Infinitely many of the corresponding  $\texttt{pre}(w)$ are palindromes, so each $M_i$ recognizes an infinite set of padded palindromes with error bound $\varepsilon$.

As for $M_i$'s runtime, note that  if $w$ is not well-padded for $f_i$, each iteration of the main loop will be terminated with probability at least $1-\varepsilon$, due to the fact that \textsc{PadCheck}$_{i,\varepsilon}$ will reject $w$ with that probability every time it is executed. The  expected number of iterations would  therefore be bounded by the constant $1/(1-\varepsilon)$. Since each iteration has polynomial expected runtime, the overall expected runtime of $M_i$ is polynomially bounded in this case.

 If $w$ is well-padded for $f_i$,
 the expected number of iterations of the main loop of $M_i$ is identical to the corresponding number for $M_{pal}$ running on $\texttt{pre}(w)$.
 To bound the runtime of $M_i$, let us consider the case where $\texttt{pre}(w)$ (whose length is 
 $\tilde f_i(n)$) is a palindrome.\footnote{Note that $M_{pal}$'s expected runtime on nonpalindromes of some length cannot exceed its expected runtime on palindromes of the same length, since the termination probability per iteration is higher for nonpalindromes.} The Acceptance Test (Figure \ref{fig:Mpal}), which is the only  cause for termination in this case, halts the machine with probability $2^{-c_{\varepsilon}(|\texttt{pre}(w)|+1)}$ in each iteration, so the expected number of iterations is in $2^{\Theta(\tilde f_i(n))}$. 
  Since $M_i$ starts every iteration  in the same quantum and classical states and with the head in the same position, we can multiply this value with the polynomially bounded expected runtime of a single iteration of the main loop of $M_i$ to obtain the overall expected runtime of $M_i$, which is  therefore in  $2^{O\left(\tilde f_i(n)\right)}$, as $\tilde f_i(n)$ dominates $\log n$.


We have proven the following  result about the family $\mathcal{L}=\Set{L_i}_{i\geq 1}$ of languages
recognized by the corresponding $M_i$s.

\begin{lemma}\label{lem:2qfacan}
    For each positive integer $i$, there exists an infinite set of padded palindromes $L_{i}$ with the following properties:
    \begin{enumerate}
        \item $L_{i}$ is associated with a padding function $f_i$ such that $\tilde f_i(n) \in 2^{\Theta\left((\log n)^{1/2^{i}}\right)}$, and
        \item $L_{i} \in \mathsf{BQTISP}(2^{O(\tilde f_i(n))},O(1))$.
\end{enumerate} 
\end{lemma}

For each $i\geq1$ and any time bound $t(n)$, we know that 
    $L_i \notin \mathsf{BPTISP}(t(n),o(\log \tilde f_i(n))$ by Lemma \ref{lem:ptmcannotpaddedpal}.
Combining this fact with Lemma \ref{lem:2qfacan}, we obtain the following fine-grained quantum advantage result:  
\begin{theorem}\label{thm:main}
    There exists a family of functions $\mathcal{F}=\Set{\tilde f_i}_{i\geq 1}$  such that for each $i\geq1$,
    \begin{enumerate}
        \item  $\tilde f_{i}(n) \in (\log n)^{\omega(1)}\cap n^{o(1)}$,   
        \item  $\tilde f_{i+1}(n) \in o(\tilde f_i(n))$, and
        \item $\mathsf{BPTISP}(2^{O(\tilde f_i(n))},o(\log \tilde f_i(n)))\subsetneq \mathsf{BQTISP}(2^{O(\tilde f_i(n))},o(\log \tilde f_i(n)))$.
    \end{enumerate}
\end{theorem}

Since $2^{O(\tilde f_i(n))}\subseteq  2^{n^{o(1)}}$ for all $i$, the infinitely many inclusions in Theorem \ref{thm:main} are the first examples of quantum computers outperforming their classical counterparts when both machines are operating within subexponential time and $\Omega(\log \log n)$ space.

\section{Concluding remarks}\label{sec:conc}


We note that the family $\mathcal{F}$ we defined in the previous section is just one of infinitely many that can be used to prove Theorem \ref{thm:main}. 2QCFAs can control a very rich collection of padding functions, both in the set $(\log n)^{\omega(1)}\cap n^{o(1)}$ that is mentioned in the theorem and also outside that range. For instance, it is easy to obtain variants of the submachine of Figure \ref{fig:padchhier} where the associated functions are in $\Theta(\log n)$, $\Theta((\log n)^2)$, $\Theta(\sqrt{n})$ \cite{S26}, $\Theta(\log \log n)$, $\Theta(\log \log \log n)$, $\Theta(\log^* n)$,   $2^{\Theta\left(\frac{\log n}{\log \log \log n}\right)}$, $2^{\Theta\left(\frac{\log n}{\log^* n}\right)}$, and so on. For any such padding function $f$, the corresponding padding control submachine can then be plugged into the template of Figure \ref{fig:baba} that generalizes Figure \ref{fig:ana} to obtain a 2QCFA for the associated set of padded palindromes.

\begin{figure}[htb!]
    \caption{General 2QCFA template for recognition of a set of padded palindromes associated with padding function $f$. $w$ denotes the input string. $\varepsilon$ is the error bound.}
    \label{fig:baba} 
    \begin{turing}[]{V}{\textit{  }}
        \titem{
            Repeat ad infinitum:}
        \ttitem{Run the polynomial-time padding control submachine (specialized for padding function $f$) on $w$.}
        \tttitem{\{This stage 
        \textit{rejects} with probability $1-\varepsilon$ if $w$ is not well-padded for $f$.\}}
        \ttitem{
            Run \textsc{PalIter}$_{\varepsilon}(\texttt{pre}(w))$.}
    \end{turing}
\end{figure}

The reason we focus on the set $(\log n)^{\omega(1)}\cap n^{o(1)}$ is that only the padding functions which are in this range allow the technique presented above to yield new results within the zone of time-space complexity that has not been explored beforehand in the context of  quantum advantage: The 2QCFA template in Figure \ref{fig:baba} yields a machine whose runtime cannot be established to be  in $2^{n^{o(1)}}$ if the padding control submachine is set to check a padding function $f$ that is not in $n^{o(1)}$ over its domain. Below the ``bottom end'', if $f(n)\in (\log n)^{O(1)}$, then the space bound $o(\log f(n))$ in the complexity class inclusion furnished by our method becomes $o(\log \log n)$, for which well-known quantum advantage results already exist \cite{AW02,R20}.

Since $2^{f(n)}$ is super-quasipolynomial for all $f(n)\in (\log n)^{\omega(1)}$,   this approach does not yield a quantum advantage result where the quantum machine has polynomial or even quasi-polynomial runtime. To see how to obtain time bounds that are ``almost'' quasi-polynomial, note that one can use the machinery of Sections \ref{sec:lo} and \ref{sec:mult} to construct and verify proof chains that certify padding functions that grow only slightly faster than any fixed power of $\log n$. For instance,  proof chains of the form $\pi_1\%\pi_2\%\pi_3\%\pi_4\%\pi_5\%\#^+$, where 
\[
|\pi_1|\in \Theta(\log n),
\]
\[
|\pi_2|\in \Theta(\log |\pi_1|),
\]
\[
|\pi_3|\in \Theta(\log |\pi_2|),
\]
\[
|\mathtt{pr}(\pi_4)|=|\pi_3|\times |\pi_2|,
\]
and
\[
|\pi_5|\in 2^{\Theta(|\mathtt{pr}(\pi_4)|)},
\]
imposing the length of the proof string $\pi_5$ to be in $(\log n)^{\Theta(\log \log \log n)}$, can handle one such padding function, and exponents with slower and  slower growth are easy to obtain by adding more and more logarithmic relationships to the chain. Also note that, since $\log \log n \in o(\log f(n))$ for all $f(n)\in (\log n)^{\omega(1)}$, the technique used for Theorem \ref{thm:main} yields 
\[\mathsf{BPTISP}(2^{O(\hat f(n))},O(\log \log n))\subsetneq \mathsf{BQTISP}(2^{O(\hat f(n))},O(\log \log n))
\]
for every padding function $f$ that is in our applicable range and for which such a polynomial-time padding-control submachine exists, where $\hat f$ is a total extension defined to emulate the asymptotic behavior of $f$, similarly to the $\tilde f_i$ of Section \ref{sec:chains}.

In a recent study of quantum complexity class hierarchies \cite{S26}, we used padding functions outside the growth range considered in this paper. Using the parlance of the present manuscript, \cite{S26} defines two language families, where, for each $i\geq 1$,
\begin{enumerate}
    \item $RPAL_i$ is a set of padded palindromes associated with a padding function in $\Theta\left(n^{1/2^i}\right)$, and
    \item $PPPAL_i$ is a set of padded palindromes associated with a padding function in $\Theta\left((\log n)^{i}\right)$.
\end{enumerate}
Two crucial theorems from \cite{S26} state the existence of 2QCFAs that recognize those languages:

\vspace{0.4em}

\noindent\textbf{Theorem 6 (of \cite{S26}).} For all $i \ge 1$, $RPAL_i \in \mathsf{BQTISP}(2^{O(n^{1/2^i})},\, O(1)).$

\vspace{0.8em}

\noindent\textbf{Theorem 11 (of \cite{S26}).} For all $i \ge 1$, $PPPAL_i \in \mathsf{BQTISP}(2^{O((\log n)^{i})},\, O(1)).$

\vspace{0.8em}

Both theorems are proven by furnishing the corresponding 2QCFA constructions. The basic idea is for the machine to perform a polynomial-time padding control procedure on the overall input string and a palindromeness check whose runtime is exponential in terms of the length of a designated prefix, as we do in Section \ref{sec:quantum}. 
However, the 2QCFAs presented in \cite{S26} perform the two controls serially, instead of following the schema of Figure \ref{fig:baba}, which interleaves   multiple executions of the padding control with a single long run of the palindromeness check. Unfortunately, this leads the constructed machines to have  expected runtime in $2^{\Theta(n)}$, but the proofs in \cite{S26} fail to arrive at this conclusion due to a mathematical misstep in the analysis. Fortunately, both proofs can be fixed by restructuring the machines as described in Section \ref{sec:Mi} (by plugging the relevant  padding control algorithm presented in \cite{S26} into the template of Figure \ref{fig:baba}), and all results of \cite{S26}, including those used in the present manuscript, stand without the need to reword any theorem statements.    

We note that the technique of Section \ref{sec:Mi} can be used to match Theorem 6 of \cite{S26}  with a counterpart\footnote{Since \cite{S26} uses a single-track alphabet,  minor modifications to the proof of Lemma \ref{lem:ptmcannotpaddedpal} would be necessary.} of our Lemma \ref{lem:ptmcannotpaddedpal} to obtain quantum advantage results where the time bounds are in $2^{o(n)}$ and the space bounds are sublogarithmic.


In conclusion, we have unconditionally proven the existence of infinitely many quantum advantage inclusions 
in the previously little-studied regime 
where the time and space budgets are in $2^{n^{o(1)}}$ and $\Omega(\log \log n)\cap o(\log n)$, respectively. A remaining open question is whether the time bounds can be improved to $n^{O(1)}$ for this range of space bounds as well. 

\section*{Acknowledgments}

The author thanks Utkan Gezer for help with the LaTeX infrastructure, and Ben Golub for giving him access to the amazingly useful AI review tool Refine, which helped improve a previous version of this manuscript. 

\section*{Declaration of AI use}

The AI tools Refine, Gemini and ChatGPT were used to provide feedback for  previous versions of this manuscript. After using these tools, the author reviewed and edited the content as needed and takes full responsibility for the content  of the published article.

\bibliographystyle{eptcs}



\bibliography{references} 

\end{document}